\documentclass[multphys,vecphys]{svmult}

\usepackage{makeidx}         
\usepackage{graphicx}        
\usepackage{multicol}        
\usepackage[bottom]{footmisc}

\usepackage{natbib}

\makeindex             

\newcommand{\hii}{\hbox{H\,\textsc{ii}}}

\newcommand{\hb}{\ensuremath{{\textrm{H}\beta}}}

\newcommand{\oiii}{[\textrm{O}\,\textsc{iii}]}
\newcommand{\msun}{\ensuremath{M_\odot}}

\newcommand{\ab}{$\sim$}
\def\asec{$''$\hspace*{-.1cm}}

\begin{document}

\title*{HST's view of the youngest massive stars in the Magellanic Clouds}


\author{M. Heydari-Malayeri\inst{1}
\and
M. R. Rosa\inst{2}
\and
V. Charmandaris\inst{3, 1}
\and
L. Deharveng\inst{4} 
\and
F.~Martins\inst{5}
\and
F. Meynadier\inst{6}
\and
D. Schaerer\inst{7}
\and
H. Zinnecker\inst{8}}

\authorrunning{Heydari-Malayeri et al.} 


\institute{LERMA, Paris Observatory, France;  
\texttt{m.heydari@obspm.fr}
\and 
ST-ECF, ESO, Germany; 
  affiliated to the Space Telescope Operations Division, RSSD, ESA;  
\texttt{mrosa@stecf.org}
\and
University of Crete, Greece;
\texttt{vassilis@physics.uoc.gr}
\and
Marseille Observatory, France;
\texttt{Lise.Deharveng@oamp.fr}
\and
MPE,  Garching bei M\"unchen, Germany;
\texttt{martins@mpe.mpg.de}
\and
LERMA, Paris Observatory, France; 
\texttt{Frederic.Meynadier@obspm.fr}
\and
Geneva Observatory, Switzerland; 
Observatoire Midi-Pyr\'{e}n\'{e}es, Laboratoire
             d`Astrophysique, Toulouse, France;    
\texttt{Daniel.Schaerer@obs.unige.ch}
\and
Potsdam Astrophysical Institute, Germany;
\texttt{hzinnecker@aip.de}}

%
\maketitle

\begin{abstract}
Accurate physical parameters of newborn
massive stars are essential ingredients to shed light on their formation,
which is still an unsolved problem. The rare class of compact \hii\
regions in the Magellanic Clouds (MCs), termed ``high-excitation blobs''
(HEBs), presents a unique opportunity to acquire this information.
 These objects (4\asec\, to 10\asec,  
\ab\,1 to 3 pc, in diameter) harbor the youngest massive stars of the OB
association/molecular cloud complexes in the MCs 
accessible through high-resolution near-IR and optical techniques.
We present a brief overview of the results obtained with {\it HST} mainly on two 
HEBs, one in the LMC (N159-5) and the other in the SMC (N81). 
\end{abstract}

\vspace*{-1.cm}

\section{Introduction}

\vspace*{-.25cm}

Massive stars ($>$ 8\,\msun) play a key role in several fields of
 astrophysics. However their formation process is still an unsolved
 problem in spite of progress, both in theory and observation, in
 recent years. 
In order to better understand their formation one needs photospheric lines to access
accurate effective temperatures ($T_{eff}$), luminosities ($L$), masses ($M$), and gravities ($g$). 
$T_{eff}$ and $L$ are crucial for placing the stars on the HR diagram and
to establish whether they are ZAMS or evolved stars. This is very important since 
massive stars evolve fast, and the most massive ones can become supergiants
while still enshrouded in clumps of  gas and dust.  
Spectral signatures are also needed to understand how and when the stellar wind
develops. Moreover, mass loss may affect the accretion rate and play an
important role in massive star formation process. Other important
issues are multiplicity and the IMF of the cluster members.
High-resolution radio continuum and IR observations yield
indispensable information about UCHII regions but do not have direct
access to newborn massive stars.

Only observations in the near-IR and optical (mainly spectroscopy) of
nascent stars can accurately provide these physical parameters.  This can be 
achieved in the visible by using the traditional methods of
spectral classification \cite{walborn06}, or by using
the more recent method of line fitting, based on atmosphere
models \cite{hillier98}, \cite{martins05}. In
the near-IR similar methodology makes use of 2\,$\mu$m spectroscopy \cite{hanson05}, 
or  profile fitting  \cite{martins05}, \cite{martins07}.

The MCs offer valuable opportunities for this study since 
they are seen face-on and have well-determined
distances. This facilitates obtaining accurate absolute magnitudes and
fluxes. Moreover, their overall extinction is low. Being metal-poor,
they provide important templates for studying star formation in
distant metal-poor galaxies which cannot be observed with comparable
spatial resolution.

\vspace*{-.5 cm}

\section{High-excitation blobs}

\vspace*{-.25 cm}

The youngest massive stars in the MCs
accessible to IR and optical observations are found in High-Excitation
Blobs (HEBs), see below for details. 
The reason for this terminology is that no spatial features could be 
distinguished with ground-based telescopes. 
This is a rare class of \hii\ regions in the MCs; so far
only 6 members have been detected in the LMC and 3 in the SMC.

For massive stars the accretion time-scale is larger than
the Kelvin-Helmholtz time-scale. This means that massive stars reach 
the main sequence while accretion is still going on. Moreover, they evolve very
fast. Therefore, obtaining the physical parameters of massive stars ``at
birth'' may be an unattainable task! Consequently, HEBs offer a compromise between stars
inside ultra-compact \hii\ regions and the exciting stars of evolved \hii\ regions.

In contrast to the typical \hii\ regions of the MCs, which are extended
structures (sizes of several arc minutes corresponding to more than
50\,pc, powered by a large number of exciting stars), HEBs are very
dense and small regions ($\sim$\,4\asec\, to 10\asec\, in diameter
corresponding to $\sim$\,1--3\,pc). They have a higher degree of
excitation (\oiii/\hb]) with respect to the typical \hii\ regions, and  
are, in general, heavily affected by local dust. 
In comparison with Galactic regions, some of HEBs are similar to classical \hii\ regions 
and some look like compact \hii\ regions \cite{martin-hernandez05}. However, 
HEBs should be considered in the context of massive star formation in the MCs.  
Compared with other compact \hii\
regions of the same sizes in the MCs, they constitute a
distinctly detached group with high excitations and luminosities. 
\cite{meynadier07}.

\vspace*{-.5 cm}

\section{LMC N159-5 (Papillon)} 

\vspace*{-.25 cm}

This prototype of HEBs is situated in the \hii\ complex N159  some 500 pc south of 30 Dor  
\cite{mhm82}. N159 is associated with one of the most important concentrations
of molecular gas in the LMC \cite{johansson98} 
and contains several signposts of ongoing star formation
(cocoon stars, IR sources, masers).   
Several physical aspects of this object were studied 
using ground-based observations in the visible, IR, and radio 
\cite{mhm85}, \cite{israel88}, \cite{israel91}, \cite{hunt94}, 
\cite{vermeij02_1}, \cite{vermeij02_2}.
 
Our high-resolution {\it HST} observations show N159 to be   
a turbulent environment typical of massive star
formation regions with stellar winds, cavities sculpted in the ionized
gas as well as shocks, absorption features, filaments, and arcs 
\cite{mhm99a}. 
 The compact \hii\ 
region is resolved for the first time into a butterfly-shaped
structure with the wings separated by \ab\,2\asec.3 (0.6 pc). Two subarcsec
features show up in the wings, a kind of ``smoke ring'' and a ``globule''
or sort of cometary structure. These features suggest a dynamic
environment, the physical details of which are still unclear. The
butterfly shape may be due to a bipolar phenomenon, but with the
present data we cannot verify. Regarding the exciting star(s), from
the \hb\ flux we expect a star of type O8 at least. The exciting star(s) is(are) 
not detected since the extinction is larger than A$_{V}$\,=\,5 mag. High-resolution
near-IR observations are necessary to penetrate into the dust
concentration. However, our  {\it JHK} VLT/ISAAC observations 
were not able to resolve the blob \cite{meynadier04}.\,Nonetheless, we 
could study the stellar populations associated with the object 
and we find that there are two main stellar populations: a
massive, young population about 3 Myr old, and an older population of
low mass stars with an age between 1 and 10 Gyr. The position of the blob  
on the color-magnitude diagram can be explained by the presence of a \ab\,50\,\msun\ 
star, affected by an extinction of A$_{V}$\,\ab\,7 mag.

\vspace*{-.5 cm}
\section{SMC N81} 
\vspace*{-.25 cm}

This object is situated in the wing of the SMC, south-east of the main bar \cite{mhm88}, 
\cite{israel88}, \cite{israel91}, \cite{vermeij02_1}, \cite{vermeij02_2}.
In contrast to  LMC N159, our {H\it ST} images reveal several stars associated with 
SMC N81 \cite{mhm99b}. Six stars are grouped in an area 2\asec\, wide. The two brightest stars, 
the main exciting sources, are only 0\asec.27 apart (0.08 pc). Two prominent dark lanes 
divide the nebula into three lobes. A magnificent curved plume of some 15\asec\, (4.5 pc) in  
length is apparently linked to an absorption hole lying towards the center of the \hii\ region. 
It is uncertain whether the plume is due to an outflow or an infall of dust.  The absorption 
hole has a size of 0\asec.25 (0.07 pc). The \hii\ region is in contact with a dense medium 
toward the west where two ridges, representing ionization/shock fronts, are present. 
This idea is corroborated by  
high-resolution ATCA interferometer observations in the radio continuum at 3 cm, which 
show compressed radio emission in that direction \cite{indebetouw04}, most probably due 
to the presence of a molecular cloud.

We also used {\it HST}/STIS to obtain UV spectra of 8 stars associated with N81
\cite{mhm02a}. Analysis of those spectra shows that all of them are 
massive stars of types O6-O8. However, they appear to be  
sub-luminous, by  \ab\,2 mag in comparison with typical O dwarfs in the MCs. 
An important property of these stars is their weakness of the stellar winds 
as evidenced by the line profiles.
 Mass loss rates as low as 10$^{-9}$\,\msun\,yr$^{-1}$, i.e. 
\ab\,2 orders of mag weaker compared with O stars of the same type \cite{martins04}.
These stars may belong to the class of Vz stars: O dwarfs lying on or
close to the ZAMS \cite{walborn92}. 
The weakness of the winds cannot be attributed to metallicity effects.
It may be due to the youth of the stars (younger stars have weaker
winds?) and if confirmed warrants further investigation.

\vspace*{-.5 cm}
\section{Other objects} 
\vspace*{-.25 cm}

Our {\it HST} observations allowed us also to resolve other HEBs for the first time. 
For example SMC N88A, which in comparison with SMC N81, is more compact 
(size \ab\,3\asec.5, \ab\,1 pc), denser, and of higher excitation, shows two components 
in interaction\,\cite{mhm99c}. The extinction  rises to more than 3.5 mag in the visible 
towards its tight core.  Such a
high extinction is unprecedented for an \hii\ region in the metal-poor
SMC. The exciting star(s) of N88A, which are certainly more massive/hotter than those of 
N81, are not detected in the optical due to the heavy extinction. 

For  LMC N11A, our {\it HST} data revealed its previously unknown nebular and stellar features, 
notably the presence of an embedded cluster of stars. Five of the stars are 
packed in an area less than 2\asec\ 
(0.5 pc), with the most luminous one being a mid O type star\cite{mhm01a}, 
\cite{parker92}.

LMC N83B displays  an impressive cavity, with an estimated age of only 
\ab\,30,000 yr, sculpted in the ionized gas by the powerful winds of a
massive star. The observations bring also to light two compact \hii\ blobs of only 
2\asec.8\,\,(0.7 pc) and 1\asec\,\,(0.3 pc) in size. The former  
 harbors the presumably hottest star of the burst and is also strongly affected by dust with
a strong extinction. 
These features plus an outstanding ridge are formed in the border zone between 
the molecular cloud and the
ionized gas possibly in a sequential process triggered by the
ionization front of an older \hii\ region.\,\cite{mhm01b}.

In addition, the two compact regions A1 and A2 lying toward the giant \hii\ region N160 
were for the first time resolved and their stellar content and morphology revealed. A1, being 
of higher excitation, is powered by a single massive star whose strong wind has created a 
surrounding bubble. A2 harbors several stars enshrouded by large quantities of dust
\cite{mhm02b}.

Apart from the papers cited above, our {\it HST} projects have resulted in   
six {\it HST}\, NASA/ESA Press Releases
(July 23, 1998; June 10, 1999;  October 5, 2000; March 28, 2001; December 19, 2001; 
September 12, 2002).

\vspace*{-.5 cm}
\section{Conclusions} 
\vspace*{-.25 cm}

Physical parameters of massive stars are needed to better understand massive star formation. 
The exciting stars of HEBs, being the youngest massive stars reachable through IR 
and optical techniques,
offer a good compromise between massive stars ``at birth'' and those in evolved \hii\ regions. 
HEBs are also particularly attractive in the context of massive star formation in the MCs. 
An unexpected and intriguing result is the weak winds of O type stars in SMC N81. 
 Are weak winds a general property of the massive stars embedded in HEBs?
 Future high-resolution near-IR observations (spectroscopy \& imaging) are needed to 
make progress. In the near future, near-IR observations with JWST and E-ELT will yield 
the photospheric signatures 
of newborn massive stars  enshrouded in A$_{V}$\,=\,100-200 mag \cite{zinnecker07} 
blobs at even earlier stages of evolution.



\printindex
\end{document}